\documentstyle[twocolumn,aps]{revtex}
\begin{document}
\draft
\title{Stripes ordering in self-stratification experiments \\ 
of binary and ternary granular mixtures}
 
\author{N.Lecocq and N.Vandewalle}

\address{GRASP, Institut de Physique B5, Universit\'e de Li\`ege, \\
B-4000 Li\`ege, Belgium.}

\date{\today}
\maketitle

\begin{abstract}
The self-stratification of binary and ternary granular mixtures has been
experimentally investigated. Ternary mixtures lead to a particular ordering of the strates which was not accounted for in former explanations. Bouncing grains are found to have an important effect on strate formation. A complementary mechanism for self-stratification of binary and ternary granular mixtures is proposed.
\end{abstract}

\pacs{45.70.Mg, 81.05.Rm, 64.75.+g}

\narrowtext

A recent experiment which has received much attention \cite{demixing} from the scientific community is the following. When a binary mixture of grains differing in size is poured between two vertical slabs, there is a global tendency for large and small grains to segregate in different regions of the pile \cite{demixing,jullien}. Additionally, a self-stratification of the mixture in alternating layers of small and
large grains is observed if large grains have a larger angle of repose than the small ones. It has been also shown \cite{herrstrat} that the phase segregation takes also place when grains of same size but with different shapes are mixed together. Strates might also appear in those conditions \cite{private_noe}.

For describing the self-stratification phenomenon, both continuous
\cite{boutreux,makse} and discrete models \cite{makse,head,nico} have been proposed up to now for binary mixtures. These models are able to reproduce the alternating layers of the different granular species. The more elaborated models consider phenomenological continuous equations which account mainly for the angles of repose of the various species, the percolation of grains in the rolling phase, and the kink formation on the pile surface. A challenge for physicists is the generalization of both experiments and models to the case of a continuous distribution of grain sizes.

The first motivation of the present work was the experimental
investigation of ternary mixtures which did not receive much attention
in earlier works. As described below, the structures of the phase
segregations in ternary mixtures raise new questions about the
self-stratification mechanisms.

A vertical Hele-Shaw (HS) cell was specially built for our purpose. The distance $e$ separating the vertical planes of the HS could be continuously adjusted. Various granular species have been used: sand, wheat semolina, poppy seeds, Fe filings. We controlled the granulometry of each species by a preliminary sifting. Important properties of each type of grains are given in Table \ref{tab:1}. Granular mixtures contained an equal volume of each species. They were poured in the HS cell with a funnel. A CCD camera with a lens 12/120mm F5.6 was placed perpendicular to the HS cell for examining the physical processes involved at the grain scale. Top views have also been taken. They allowed us to observe the dynamics and composition of the upper surface during the flow. Precise video imaging is necessary for studying the evolution of granular piles. Our results are reproducible.

First, let us consider the case of ternary mixtures being composed of small ($S$), medium ($M$) and large ($L$) grains. Fig.\ref{fig1} presents three pictures of ternary piles in the HS cell. Two different mixtures were used. In each case, size segregation is clearly observed. Indeed, small grains have a global tendency to locate in the center of the pile while large grains are located in the tail. The medium-sized grains are dispersed in between the center and the tail. In addition, the self-stratification phenomenon is present. Three different patterns are obtained. In the first pile, a $...SMLSMLSML...$ ordering is displayed as found in \cite{demixing}. The second pile is characterised by a $...SMLMSMLM...$ ordering. The third pile is quite similar to the
first, as far as wide stripes are concerned. However, the ordering seems to be more complex, especially in the tail: a thin stripe of small grains is visible in the center of wide stripes made of large grains. These different orderings require special attention and the presence of the thin stripe cannot be explained with conventional models for self-stratification. Another important observation is that
for some mixtures, a difference in inclinaison angle appears between the different stripes (Fig.\ref{fig1}$b$). This difference of stripe angles, the particular ordering and the occurrence of a thin additional stripe indicates that quite different mechanisms are present during the heap formation.

In the case of binary mixtures, Koeppe {\it et al.} \cite{koeppe} have reported some anomalous stripe pattern that they called ``pairing". Indeed, stripes of the same grains seemed to be formed by pairs in particular experimental conditions. We made careful experiments with binary mixtures and we observed again the above-mentioned thin stripes (see Fig.\ref{fig2}). We suggest that thin stripes can be at the origin of the previously reported ``pairing". Similarily to ternary mixtures, the successive stripes may have different angles.

Self-stratification for binary mixtures of grains differing in size has been explained \cite{demixing,makse2} by the combination of two main processes. They are illustrated in Fig.\ref{fig3}. The first process is the size segregation mechanism. There is a global tendency of large grains to roll farther downhill than the small ones due to their larger mass/inertia. This effect is strongly increased by the percolation of
small grains through the gap between large grains in the flow. Large grains will be on the top of small grains, in the upper part of the moving layer. Small grains form a relatively smooth surface, on which large grains roll easily. Thus the moving layer is characterized by a strong velocity gradient such that the upper large grains are the fastest. As a result, large grains also locate in the front of the avalanche while small grains are confined in the lower part of the tail of the moving layer. Large grains will travel farther than small grains. The second process coming into play is the so-called kink mechanism. When the head of an avalanche reaches the end of the pile, grains will stop moving due to the horizontal slope of the basis. If the flux is not too important down the slope, grains in the tail of the avalanche will come to rest on the wall just formed by preceding grains. This wall is the kink and it moves up the slope as more grains arrive on it. This process is similar to the formation and backing of traffic jams. A pair of layers is formed through the kink mechanism with the small grains in the lower layer and the large grains in the top layer, starting at the basis of the pile. Since small grains were located in the tail of the avalanche, the wide stripe of small grains does not extend to the basis of the pile. The resulting surface involves an efficient capture for small grains during the next avalanche.

This explanation can be extended to ternary mixtures. It accounts for patterns in Fig.\ref{fig1}$ab$. The $...SMLMSMLM...$ ordering reflects the dynamics and composition of the lower layer of the avalanche: large grains in the head, followed by medium sized grains, themselves followed by small grains. So when an avalanche flow down on top of the surface formed by the preceeding kink, after the head of large grains has passed, medium grains will pass and some will be captured before small grains pass. A very close examination of the heap displayed in Fig.\ref{fig1}$a$ reveals that there are indeed some medium sized grains caught below the stripes of small grains in the lower part of the heap. On the other hand, in the upper part of the heap, the $...SMLSMLSML...$ ordering is visible because there was not enough time for medium grains to move in front of small grains in the lower part of the moving layer. Thus, small grains were the first to be captured during the avalanche in the upper part of the pile. These dynamical processes depend strongly on the incoming flux of grains and the relative velocities of different granular species. However, the above mechanism for self-stratification cannot explain the thin layers of small grains observed in Fig.\ref{fig1}$c$ and in Fig.\ref{fig2}.

The complementary mechanism we propose is illustrated in Fig.\ref{fig3}. Considering again a binary mixture, the first steps are identical to previous explanation. The kink, however, does not catch all moving grains. Indeed, in any granular flow, various transport mechanisms take place. From bottom to top in the moving layer, one encounters: combined movement of grains, motion of individual grains with friction and collisions, and above the rolling-slipping phase, a layer with boucing grains. Typically, the thickness of the rolling-slipping phase is about five to ten large grain diameters in our mixtures. The bouncing grains layer is characterised by a higher velocity and a much smaller density. Indeed grains does not touch each other permanently like in the 
rolling-slipping phase. We found that bouncing grains can jump as far as 2 cm and then rebounce or be captured. We stress that most bouncing grains found at the bottom of the pile have bounce all their way down. Both kinds of grains can bounce. The efficiency of this mechanism depends on resilience of the grains. We observed that a non-negligible amount of those bouncing grains can flow down the slope over the kink (see Fig.\ref{fig4}). We stress that bouncing grains are those involved in the formation of the additional thin stripe observed in the center of the wide stripe made of large grains.

Fig.\ref{fig5} shows also the evolution of a binary pile. The first picture was taken as the kink was moving up. The thickness of the kink is about the same that the thickness of the rolling layer. As a consequence, the kink moves fast upward. The second picture shows the pile just after the kink has passed. The third picture was taken just before the next avalanche comes down, and the last picture shows the pile as the avalanche is flowing.

It is clearly visible that the thin stripe of small grains appears mainly during the time interval separating pictures $b$ and $d$, that is between the passage of the kink and the next avalanche. In addition, a close examination shows that this thin stripe is composed of both kinds of grains, in opposition to wider stripes where the granular species are quasi-pure. In the case of pile 1$c$, close look reveals that small, medium and large sized grains are involved in the thin stripes.

The reason these strates are thiner is that there are much less grains involved -- only those that are not immediatelly stopped by the kink. When a small boucing grain is captured by the pile, it will fall through the gaps left by the large grains forming the static surface. This is why the stripes are formed of both kinds of grains. This also explains the fact that the thin stripe is observed to form below the static surface, about one large grain deep. Thin stripes have the same inclinaison angle as the surface left by the kink. The difference in angle for thin stripes and wide stripes is observed in Figure 2$b$ ($\Delta\theta\approx 3^\circ$). It indicates that the kink builds at a greater angle than the angle of the avalanche (avalanches erode mainly the upper part of the pile). This can be related to another observation we made when pouring granular materials through the funnel. We noticed 
that the jamming limit diameter was lower in the case of mixture made of large and small grains than in the case of large grains only. Thus the presence of small grains seems to reduce the arching properties of large grains. We do not know if this is a general feature.

Several causes can inhibe or hide the phenomena of thin stripe formation. In the case of Fig.\ref{fig1} $a$ and $c$, the only difference is the funnel used for pouring the grains. As a consequence, not only the flux of grains was lower in the case of pile $a$ but also, and this is related, the initial velocity of grains seemed to be lower in $a$. The main effect is a lowering of the number of bouncing grains. The kink may become perfectly efficient, i.e. it may completely stop all moving grains it encounters. Then the thin layer will not form. Indeed we observed no boucing grains flowing over the kink during heap
formation of $a$. Next, if the avalanches are important, they can erase by erosion the very thin layer that forms the top of the pile.

All the mixtures we used are characterised by a large aspect ratio ($>2$, up to 10) and by a larger density for small grains. Koeppe {\it et al.} \cite{koeppe} used a mixture of sand and sugar that corresponds to these characteristics when they obtained their ``pairing''. The difference in size causes percolation to be efficient and might favorise an important velocity gradient in the moving layer. These make self-stratification particularly pronounced. High velocity in the top of the rolling-slipping layer favorise the transport of bouncing grains. The large density of small grains allows them to bounce as far as large grains do. Indeed, in the case of grains all of the same density, small grains are much lighter than large grains. They would have much less kinetic energy and so they would travel less far than large grains. 

The number of grains involved in the thin stripe, depending on the position along the slope, could be computed from the time interval between the passage of the kink and next avalanche, and the number of bouncing grains at that point, given that one knows the probability of capture by unit lenght for those grains by the pile. We observed that the thin stripe of small grains do not always presents the same 
profile, in particular, for some mixtures with narrower density difference between small and large grains, thin stripes do not extend to the basis of the pile. Taking the problem backward, this could provide an efficient way to study the transport properties of bouncing mechanism.

In summary, we have investigated both ternary and binary granular mixtures in vertical HS cells and have found particular patterns. We observed the formation of an additional thin stripe which we relate to the previously reported ``pairing'' \cite{koeppe}. This cannot be fully explained by former self-stratification models. We have proposed a complementary mechanism which is compatible with observations of binary/ternary patterns. The mechanism is based on the presence of bouncing grains on top of the rolling-slipping layer. Size ratio and denstity difference between small and large grains seems to be decisive properties of mixtures for thin stripe forming. In particular, denser small grains can bounce as far as large grains. Two important experimental parameters were incoming flux and initial velocity of grains. More work on the subject is needed. 

NV thanks the FNRS (Brussels, Belgium) for financial support. This work is also supported by the Belgian Royal Academy of Sciences through the
Ochs-Lefebvre prize. Valuable discussion with M.Ausloos, J.Rajchenbach,
E.Cl\'ement, S.Galam are acknowledged.

\begin{table}
\centering
\begin{tabular}{lrcc}
material          & size          & repose angle & density\\
\hline
sand 1 ($e=6$ mm) & 0.25-0.315 mm & 37$^{\circ}\pm2$ & 2.6 g/cm$^3$\\
sand 1 ($e=2$ mm) & 0.25-0.315 mm & 41$^{\circ}\pm1$ & 2.6 g/cm$^3$\\
sand 2            & 0.2-0.25 mm   & 42$^{\circ}\pm1$ & 2.5 g/cm$^3$\\
sand 3            & 0.25 mm       & 35$^{\circ}\pm1$ & 2.8 g/cm$^3$\\
Fe filings ($e=6$ mm)& $<0.1$ mm  & 35$^{\circ}\pm2$ & 5.2 g/cm$^3$\\
Fe filings ($e=2$ mm)& $<0.1$ mm  & 42$^{\circ}\pm1$ & 5.2 g/cm$^3$\\
wheat semolina    & 0.8 mm        & 41$^{\circ}\pm1$ & 1.3 g/cm$^3$\\
poppy seeds       & 1 mm          & 39$^{\circ}\pm1$ & 1.1 g/cm$^3$\\
\end{tabular}
\caption{Properties of the grains we used.}\label{tab:1}
\end{table}

\begin{figure}
\caption{Three different patterns observed in ternary granular mixtures. The mixtures are composed of small ($S$), medium ($M$) and large ($L$) grains. (a) A pile exhibiting a $...SMLSMLSML...$ type of stripes ordering ($S$ in black, $M$ in grey and $L$ in white). Close look at the tails of the stripes reveals a $...SMLMSMLM...$ ordering. (b) $...SMLMSMLM...$ ordering with clear $M$ grains and dark $S$ and $L$ grains. A difference in angle between $L$ stripes and $S$ stripes ($S$ in the upper part of the heap) is emphasised. (c) Additional thin stripes are visible mainly in the tail of the pile. The same mixture was used in $a$ and $c$ (Fe filings ($S$), sand 1 ($M$) and wheat semolina ($L$)) with the same plate separation (6 mm). Only the funnel used was different (2 cm$^3$/s in $a$ and 2.5 cm$^3$/s in $c$, with faster grains). The mixture used in $b$ was composed of Fe filings ($S$), sand 2 ($M$) and sand 1 ($L$) with a plate separation of $e=2$ mm. Images $a$ and $c$ have been processed by computer for better contrast between the three different granular species.} \label{fig1}
\end{figure}

\begin{figure}
\caption{Additional stripe of small grains in binary mixtures with 
$e=6$ mm plate separation. (a) A binary granular mixtures of sand 3 ($S$) and poppy seeds ($L$) gives rise to pairs of $L$ strates. Another way to describe the pattern is to say that a thin $S$ stripe is in the center of each $L$ stripe. (b) Sand 1 ($S$) and wheat semolina ($L$). Notice the difference in angle between the wide $S$ layers in the upper part of the pile and the thin $S$ layers of the lower part.} \label{fig2} 
\end{figure}

\begin{figure}
\caption{Schematic representation of self-stratification mechanisms described in the text for a flow of binary mixtures in a HS cell. Small grains are in dark grey and large grains are in light grey. (left column) Three stages of the formation of pairs of strates in earlier explanations.
(right column) Three stages of the formation of strates in our
model.}
\label{fig3}
\end{figure}

\begin{figure}
\caption{Upper part of pile 1c during formation. Bouncing grains are flowing over the kink.} 
\label{fig4}
\end{figure}

\begin{figure}
\caption{Apparition of a thin stripe of small grains in a 
binary mixture (Fe filings ($S$) and wheat semolina ($L$), plate separation $e=6$ mm). (a) The kink is pointed by an arrow. (b) Just after the kink has passed. (c) Just before the next avalanche. The thin $S$ stripe has appeared. (d) The head of the next avalanche is marked by an arrow.}
\label{fig5}
\end{figure}

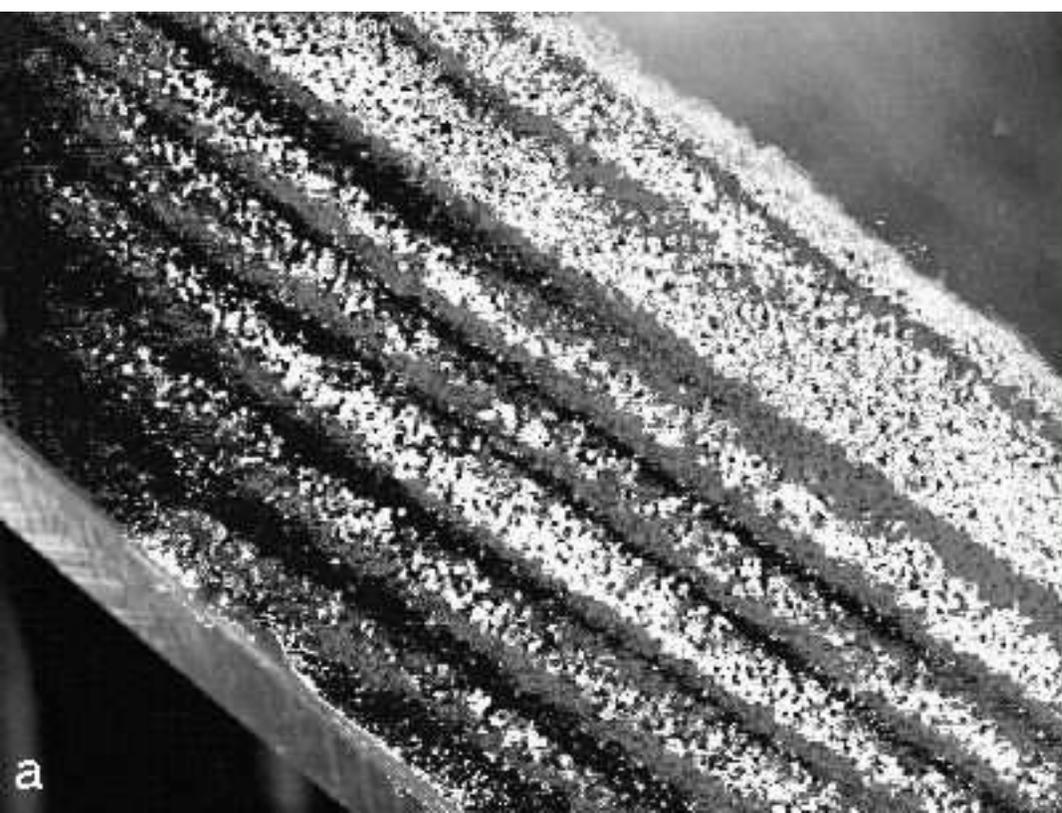
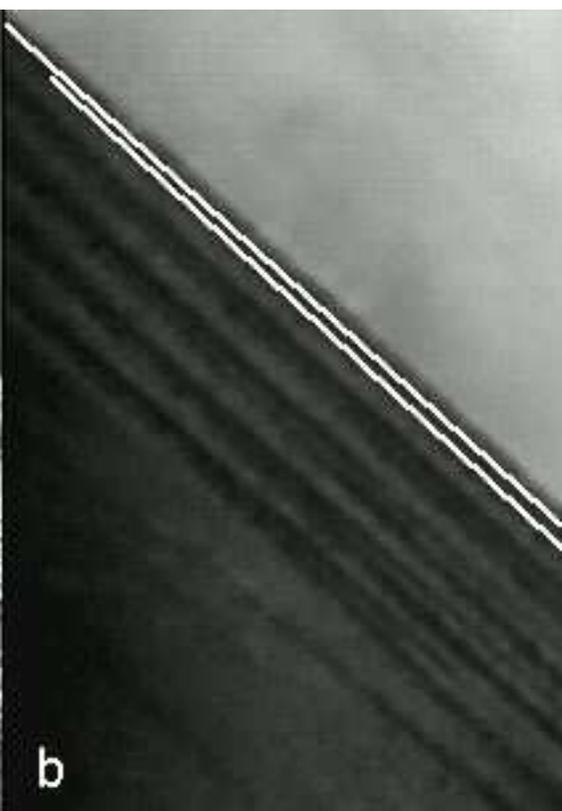
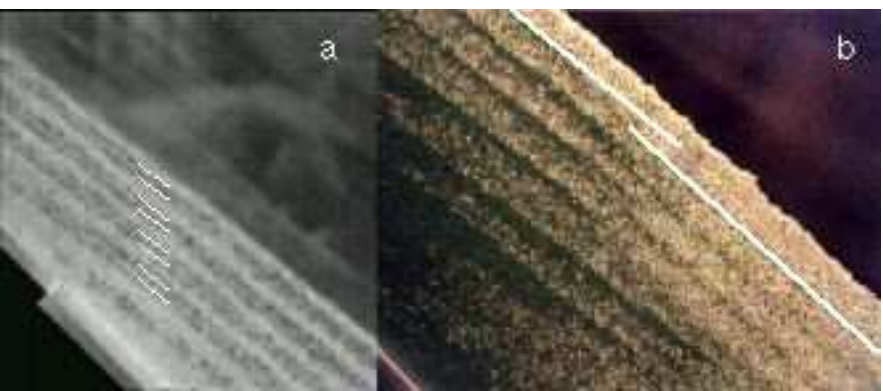
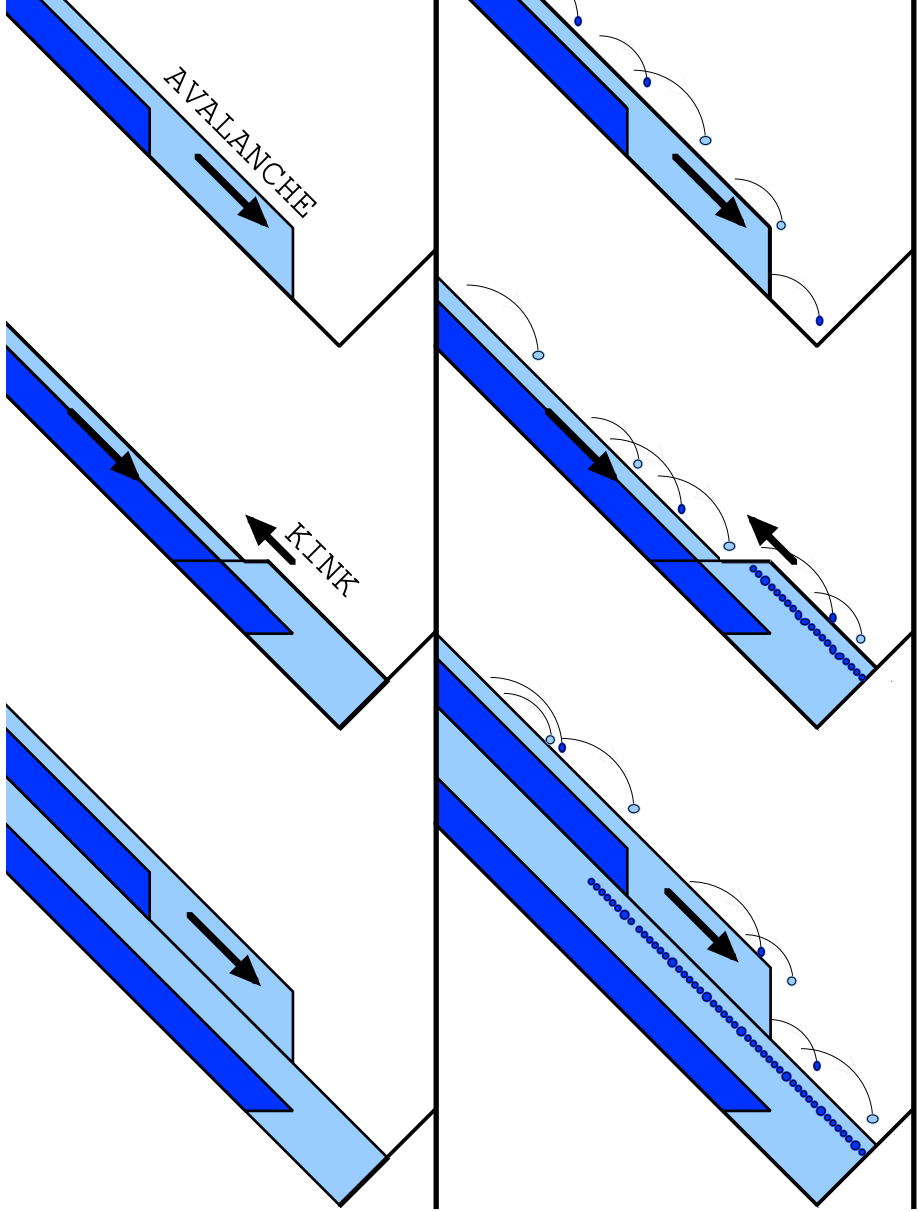
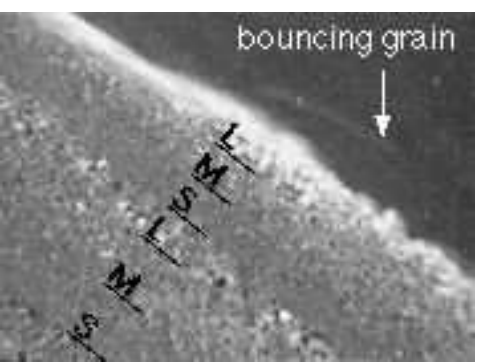
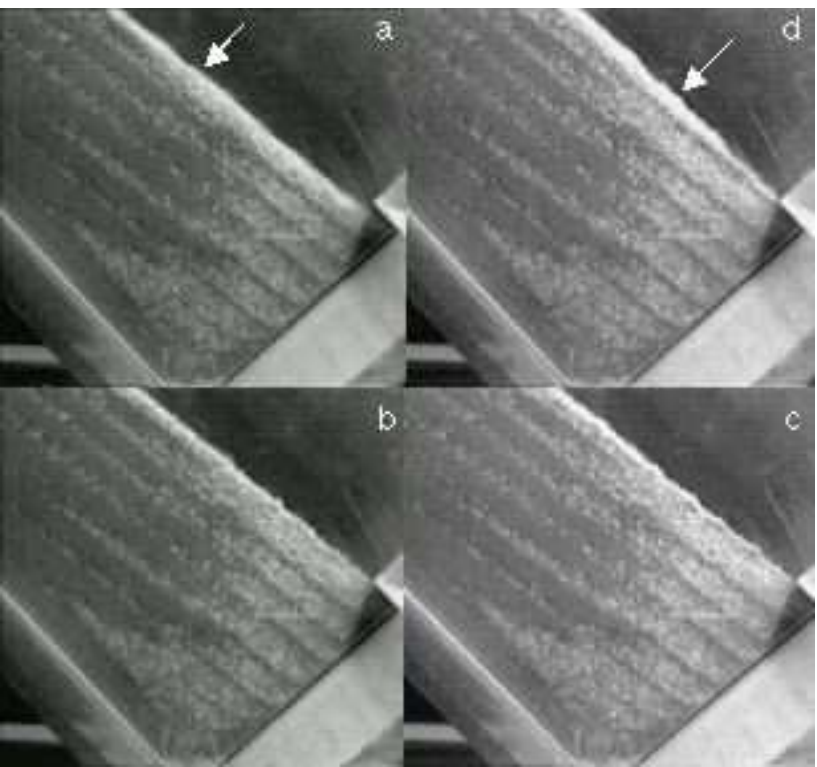
\end{document}